# A hybrid bio-organic interface for neuronal photo-activation

Maria Rosa Antognazza[1,°], Diego Ghezzi[2,°], Marco Dal Maschio[2,°], Erica Lanzarini[1,3], Fabio Benfenati[2, °,*] & Guglielmo Lanzani[1,3, °,*]

[1]*Center for Nanoscience and Technology of IIT@POLIMI, via Pascoli 70/3, 20133 Milano, Italy.*

[2]*Department of Neuroscience and Neurotechnologies, Italian Institute of Technology, Via Morego 30, 16163 Genova, Italy.*

[3]*Politecnico di Milano, Physics Dept., P.zza L. da Vinci 32, 20133 Milano, Italy.*

[°]*These authors contributed equally to this work*

[*]*Corresponding authors. F.B.: Tel.: +39 010 71781 434; fax: +39 010 720321; e-mail address: fabio.benfenati@iit.it; G.L.: Tel.: +39 02 2399 6166; fax: +39 02 2399 6126; e-mail address: guglielmo.lanzani@iit.it*

**Interfacing artificial functional materials and living neuronal tissues is at the forefront of bio-nano-technology[1]. Attempts have been so far based onto micro-scale processing of metals and inorganic semiconductors as electrodes[2] or photo-active layers in biased devices[3,4]. More recently, also nanomaterials properties have been investigated[5]. In spite of extensive research however, the communication between biological tissues and artificial sensors is still a challenge. Constraints consist in the complexity of the fabrication processes (i.e. metal and semiconductor lithography), the mechanical properties (e.g. flexibility and mechanical invasiveness) and chemical influence (e.g. inflammatory reactions). In addition, electrodes have fixed geometries that limit the location in space of the stimulus and often electrical currents are detrimental for the overall system.**
**To this respect organic soft matter offers a chance in terms of biological affinity[6-8] and mechanical properties. In particular conjugated polymers have appealing optoelectronic features which could lead to a new generation of neuronal communication and photo-manipulation techniques. So far conjugated polymers have being only tested as coatings of electrodes for neuronal activity recording[9-13]. Here we report an up-scale of their use: the successful interfacing of an organic semiconductor to a network of cultured primary neurons, through optical excitation. This allows to a new paradigm for the optical stimulation of neurons which could have important implications for the development of an artificial retina based on organic photodetectors.**

The organic photodiode active layer is the prototype material for photovoltaic applications[14], namely regio-regular poly(3-hexylthiophene- 2,5-diyl) with phenyl-C61-butyric-acid-methyl ester ([60], (rr-P3HT:PCBM) (see Fig. 1a).

The device, whose structure is schematically displayed in Fig. 1b, is realized through a multi-stage process: first, the active polymer film is deposited by a standard spin-coating deposition technique, onto a glass substrate pre-coated with Indium-Tin Oxide (ITO), which works as the anode of the photodetector. Second, the organic blend is annealed at 120 °C for two hours. The thermal treatment plays a double role: (i) it improves the morphology of the polymeric film, enhancing the efficiency of charge photogeneration[15], and (ii) it prepares the film for subsequent cell culture, by removing all the possible residual solvents (e.g. acetone, methyl alcohol, chlorobenzene), highly toxic for the biological systems, and sterilizes it, with the thermal elimination of biological pollutants. Third, the polymer layer is covered by a classical adhesion molecule, poly-L-lisine (PLL), deposited by casting, and primary rat embryonic hippocampal neurons are finally seeded and grown on top of it.

A successful device demands specific requirements to be fulfilled by the organic semiconductor, namely: (i) the organic semiconductor must stand a sterilization procedure, (ii) the organic semiconductor must survive and preserve its characteristics once immersed in culturing medium, (iii) cell growth and survival must be demonstrated on top of our organic active layer and (iv) proper functionalities of the semiconducting layer and of the neuronal network have to be demonstrated.

To address issues (i) and (ii) we preliminarily investigated prototypes not including neurons but undergoing the same fabrication steps[16]. We used hybrid solid-liquid devices using culturing medium, recording extracellular saline solution (Ringer) or aqueous solution of sodium chloride as electrolytic cathodes; the latter was used as a simpler and proper model system for understanding the hybrid photodetector operation's principle, considering that NaCl represents the main ionic content of any culturing medium.

Data in Fig. 1c clearly demonstrate that the photosensing capabilities of the organic polymer are fully preserved after thermal sterilization, casting of PLL and direct contact with saline solutions for 12 days. We verified that the hybrid cell can work *continuously* for more than 30 hours, with negligible reduction of its efficiency, and lasts for more than one month.

Fig. 2a-b shows optical microscope images of primary neurons grown onto either PLL-treated ITO/rr-P3HT:PCBM devices or control substrates. Their adhesion to the substrate and growth on ITO/rr-P3HT:PCBM devices were comparable to the control conditions. By adopting a standard patch-clamp technique, we recorded spontaneous activity (Fig. 2c) and action potentials induced by current injection into cultured neurons (Fig. 2d). These results demonstrate that *organic layers work effectively as cell-culturing substrates*, without affecting the adhesion and growing properties of primary neurons and that cultured neurons preserve their fundamental physiological characteristics in the presence of the conducting polymer.

Next, we tested the effectiveness of our interface structure as an inter-communication device by evaluating if light excitation and consequent current generation affect neuronal activity. To this purpose, electrophysiological experiments under photo-stimulation were performed in *photovoltaic mode* ($V_b = 0$ V).

We found that there is a deterministic correspondence between photo-stimulation of the organic semiconductor and neuronal activation in both current-clamp (Fig. 3c) and voltage-clamp (Fig. 3d) experiments. This establishes a communication protocol which translates photons into neuron activation and signalling.

The photo-stimulation mechanism here performed seems well compatible with the neuronal physiological condition. For instance we directly monitored the pH changes in the extracellular solution in close proximity of the recorded cells (Fig. 3b) both in the presence of photo-stimuli and in control experiments of the same duration in which photo-stimuli were not delivered. We found that the local pH change did not significantly drift over 0.02 pH units, which are anyway unlikely to affect neuronal function.

As a check experiment, neurons cultured directly onto glass/ITO substrates did not show any response to illumination, suggesting that neither thermal mechanisms nor direct light gating of neurons can be considered responsible for the photo-induced stimulation of the system[17,18]. This is shown by the Post Stimulus Time Histograms (PSTH) resulting from five consecutive light pulses (Fig. 3d), in which the light induced activation of neurons cultured on ITO/rrP3HT:PCBM devices (black trace) is virtually absent from neurons cultured onto control glass substrates (red trace).

As explained in Fig. 1d, the physical mechanism underlying the photostimulation process could be either a pure resistive ($R_i$) behaviour, or a pure capacitive ($C_i$) behaviour, or a combination of the two. In the first scenario, following polymer photo-excitation and photocurrent generation, the charge migration from the polymer to the electrolyte gives rise to Faradaic currents; subsequently, an ionic unbalance is created in the extracellular medium. Given the high resting chloride anions conductance of neurons, the decrease of $Cl^-$ could drive a gradient outflow of chloride anions from the neuron (inward current), giving rise to membrane depolarisation and subsequent firing. This scenario is usually taken into consideration whenever an external bias is applied, stronger than the relevant electrochemical potentials and high enough to enable water hydrolysis (in our case higher than 3.4V). Note that Faradaic processes resulting from reduction-oxidation electrochemical reaction between species in solution and electrodes might induce both electrodes damage and cell degradation, and for this reason should be avoided.

The photo-stimulation can also trigger a pure-capacitive current, giving rise to the generation of two oppositely charged Helmholtz layers at the electrolyte/polymer and neuronal membrane/electrolyte interfaces, without charge transfer processes between the polymer and the electrolyte[19]. The charge displacement in the extracellular space depolarizes the neuronal membrane and elicits the action potential.

Considering that (i) we work in photovoltaic mode, (ii) the recorded photocurrent is low (in the order of few hundreds of pA) and (iii) during the stimulation experiments we do not observe any adverse effect (neither modification of the electrodes, nor of the neuronal culture), the capacitive coupling seems to be the most likely mechanism of stimulation.

A Faradaic component of the current can not be completely excluded, but in our case should be negligible; if present, a significant increase in the concentration of hydroxide ions is expected, contrary to our experimental evidence.

In conclusion, we demonstrate *a new communication protocol between organic semiconductors and neuronal cells, showing photo-stimulation of neuronal activity*. Our approach has advantages respect to similar attempts based on biased devices[3,4]. In contrast with metal or silicon interfaces, the proposed interface works without any externally applied electric field and with minimal heat dissipation, favourably addressing the thermal issues, extremely relevant in an efficient biological interface. It was demonstrated that by using organic semiconductors one can closely reproduce the human retina colour response functions[20]: this work suggests that π-conjugated materials can be taken into consideration for a new generation of bio-mimetic artificial retinal prostheses. Organic technology is characterized by simple and cheap fabrication techniques; existing deposition methods, such as ink jet printing, allow the realization of a variety of geometrical patterns with different active areas, up to few square micrometers, thus offering the possibility to specifically target selected groups of cells. The use of soft matter provides some advantages in terms of mechanical properties, since it allows to produce light, thin and flexible devices, better suited for implantation within a biological environment.

In perspective, our approach is a simpler alternative to the existing and widely used neuron optogenetic photo-stimulation techniques[21] and represents a new tool for neural active interfacing.

**Methods**

**Semiconducting polymers.** Both rr-P3HT and PCBM were supplied by Sigma-Aldrich and used without any further purification. rr-P3HT has a regio-regularity of 99.5 % and molecular weight of 17500 g mol$^{-1}$. An accurate cleaning of the substrate was required: the substrate was rinsed in an ultrasonic bath with, subsequently, a specific tension-active agent in water solution (3%), deionised water, pure acetone and isopropyl alcohol. An oxygen plasma cleaning of the substrate completed the process. 1,2-Chlorobenzene solutions of P3HT and PCBM were prepared separately, with a concentration of 7,5 g l$^{-1}$, and then mixed together (1:1 volume ratio) using a magnetic stirrer. The solution was then heated at 50 °C, stirred and finally deposited on the ITO-covered (thickness 0.4mm) glass substrate, previously heated, by spin-coating. Spinning parameters (rpm, speed, rotation duration) were carefully selected in order to obtain suitable optical quality and film thickness (~150 nm). After deposition, organic layers were annealed and properly sterilized by heating at 120 °C for 2 hrs. Control substrates (ITO-covered glass substrates) were properly sterilized in the same way.

**Photocurrent measurement.** In the modulated photocurrent spectroscopy, a 30 W halogen tungsten lamp served as a source of white light, filtered through a monochromator before being focused onto the photodetector. The light was chopped by a mechanical chopper, whose reference signal was fed to a lock-in amplifier, with detection at the chop frequency (270 Hz). All measurements were performed at room temperature and without applying bias. The hybrid device had a vertical geometry, with an ITO layer that served as an anode, the active polymer deposited on top of it, and an aqueous NaCl (200 mM) solution used as the electrolytic, liquid cathode. Other solutions tested for the cathode included: Minimum Essential Medium (MEM), recording extracellular solution (Ringer, see below) and several aqueous solutions with various concentrations of other salts, like NaI or NaBr. The photocurrent was extracted and measured in the external circuit through a reference electrode (gold), immersed in the electrolyte.

**Cell Culture.** Primary cultures of hippocampal neurons were prepared from embryonic day 18 rat embryos (Charles River, Calco, Italy). Briefly, hippocampi were dissociated by a 15 min incubation with 0.25% trypsin at 37 °C, and cells were plated at a density of 450-550 cells/mm$^2$ on poly-L-lysine (0.1 mg/ml)-treated organic layers and control substrates in Neurobasal (Invitrogen, San Giuliano Milanese, Italy) supplemented with 10% horse serum (Hyclone, Logan, UT). After allowing neurons to adhere to the surface for 3-4 hrs, cells were cultured in serum-free Neurobasal supplemented with 2 % B27 (Invitrogen) and 2 mM glutamine (Invitrogen).

**Photo-stimulation and Electrophysiology.** Electrophysiology was performed on a set-up based on a Nikon FN1 upright microscope (Nikon Instruments, Calenzano, Italy). Differential Interference Contrast (DIC) images were taken by a C91006 CCD Camera (Hamamatsu Photonics Italia, Arese, Italy). Using a custom LabView application (National Instruments, Milano, Italy), a specific Region of Interest (ROI) surrounding the soma of a selected neuron was drawn and the resulting binary image was used to control a Digital Micromirror Device (Texas Instruments, Dallas, TX, USA). The DMD was in turn used to shape a laser beam (Cobolt, Solna, Sweden) resulting in a selective photo-stimulation limited to the defined ROI. Laser pulses (532 nm, 10 mW/mm$^2$) were timed using a mechanical shutter (Uniblitz, Rochester, NY, USA) and delivered to the sample via a 40x/0.8NA water immersion objective (Nikon Instruments). Whole-cell patch-clamp recordings were performed at room temperature by employing patch-pipettes (4-6 MΩ), under GΩ patch sealing, using an Axopatch 200B (Axon Instruments, Foster City, CA, USA). The recording extracellular solution (Ringer) contained (mM): NaCl 135, KCl 5.4, $MgCl_2$ 1, $CaCl_2$ 1.8, HEPES 5, glucose 10, pH 7.4 adjusted with NaOH. The intracellular solution contained (mM): KCl 140, HEPES 5, EGTA 5, $MgCl_2$ 2, pH 7.35 adjusted with KOH. Responses were amplified, digitized at 20 kHz and stored with pCLAMP 10 (Axon Instruments). Further analyses were done using pCLAMP 10 and MATLAB (The MathWorks, Torino, Italy).

**Acknowledgements**

This work was supported by grants from Compagnia di San Paolo – Torino (to F.B.), thelethon Italy GGP09134 (to F.B.) and Ministero dell'Univerità e della Ricerca PRIN Project (to F.B.).

**Author contributions**

All authors contributed equally to this work. G.L. and F.B. conceived the experiment. M.R.A. and E.L. prepared the polymer films and performed the photocurrent experiments. D.G. and M.D.M. prepared primary neuronal cultures and performed electrophysiological experiments. M.R.A. wrote the main paper. All authors discussed the results and commented on the manuscript at all stages.

**Additional information**

The authors declare no competing financial interests. Reprints and permission information is available online at http://npg.nature.com/reprintsandpermissions/. Correspondence and requests for materials should be addressed to F.B. and G.L.

**Figure captions**

**Figure 1 | Solid-liquid organic photodetector: the operation principle.** (a) Chemical structures of rr-P3HT and PCBM. (b) Scheme of the photo-sensing interface, with the neuronal network grown on top of the polymer active layer during patch-clamp recordings. (c) Spectral responsivity of the systems ITO/rr-P3HT:PCBM/NaCl/Gold and ITO/rr-P3HT:PCBM/Ringer/Gold, with the saline solutions working as ionic cathodes, recorded after 12 days of immersion. (d) Possible models of the polymer/electrolyte and electrolyte/neuron interfaces, where $C_i$ and $R_i$ represent, respectively, the capacitance and the resistance of the double layer at the interface of electrolyte and polymer. $V_b$ represents the bias voltage traditionally applied for the cell stimulation (in our case $V_b = 0$) and $R_s$ the electrolyte resistance.

**Figure 2 | Bio-organic active interface.** Hippocampal neurons cultured after 12 DIV on either poly-L-lysine treated ITO/rr-P3HT:PCBM devices (a) or control glass substrates covered only with ITO and poly-L-lysine (b). Scale bars, 50 μm. (c) Whole-cell recording showing spontaneous activity in a neuron cultured over the ITO/rr-P3HT:PCBM device. (d) Single action potential induced by current pulse injection into the same neuron in (c).

**Figure 3 | Optical stimulation of neurons cultured onto an ITO/rr-P3HT:PCBM device.** (a) The optical stimulation paradigm includes the localisation of the stimulus in a region surrounding the patched neuron (scale bar, 10 μm). (b) On-line monitoring of the pH of the extracellular solution during photo-stimulation experiments (black) or

without photo-stimulation (red). Points are *means ± S.D.*, *n = 4*. (c) Action potential generation in response to photo-stimulation pulses of different duration. A 100 ms stimulus generates a single action potential (left), whereas a 500 ms stimulus generates a burst. (d) Voltage-Clamp ($V_{hold}$ = -70 mV) recording upon light illumination (mean of *n = 5* consecutive trials). (e) Post-Stimulus Time Histogram (PSTH) on neurons cultured on ITO/rr-P3HT:PCBM devices (black) and control substrates (red) subjected to a series of light pulses (1Hz, 100ms) (*n = 5* cells).

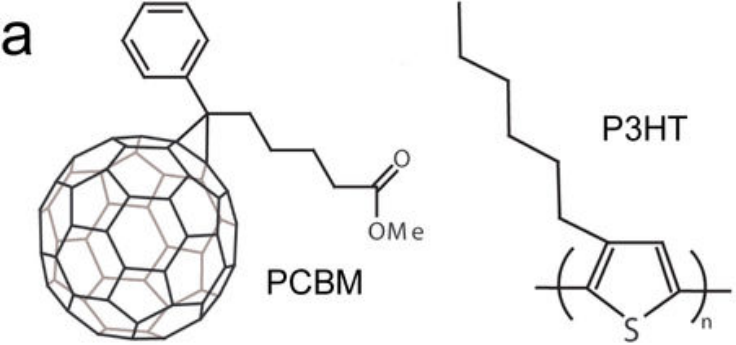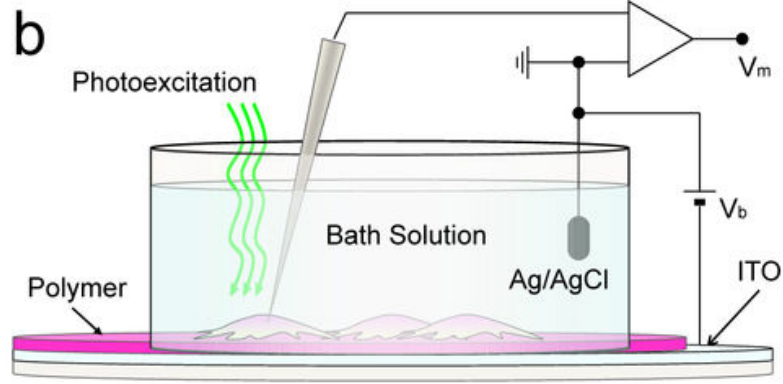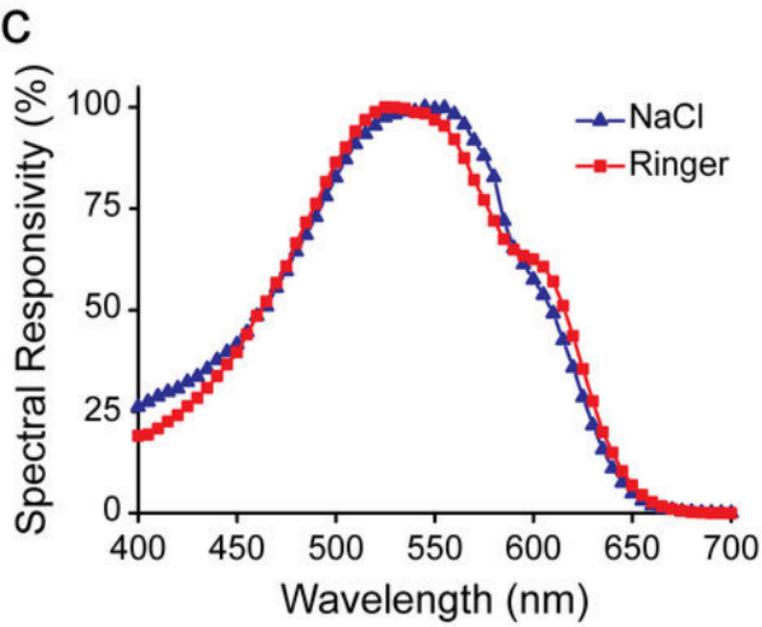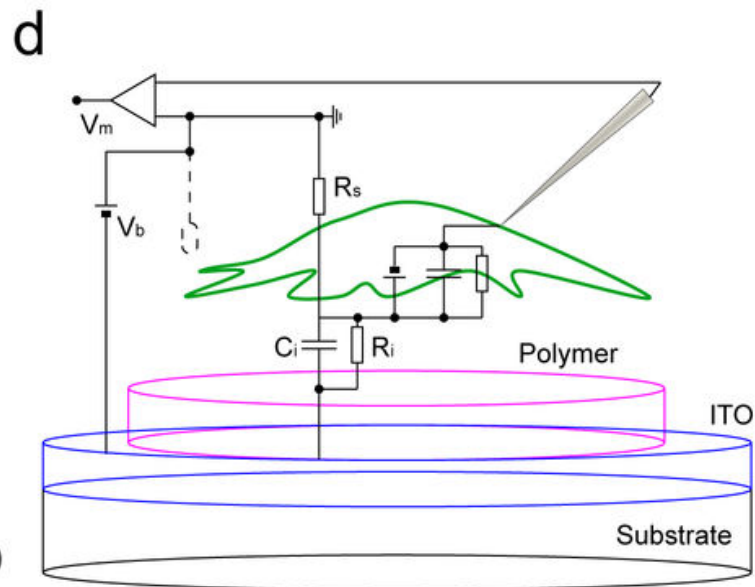

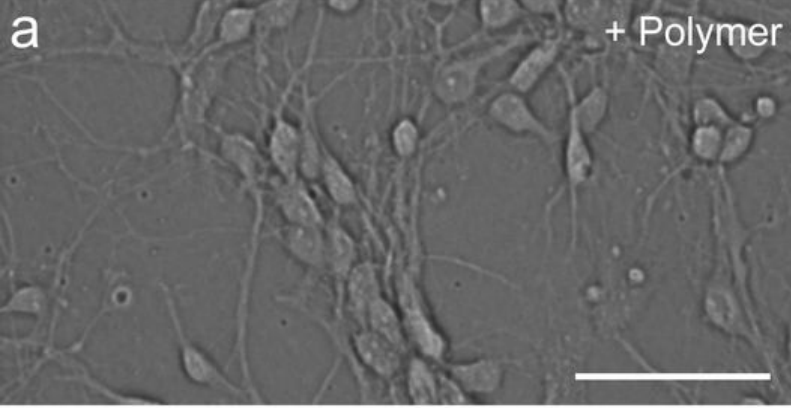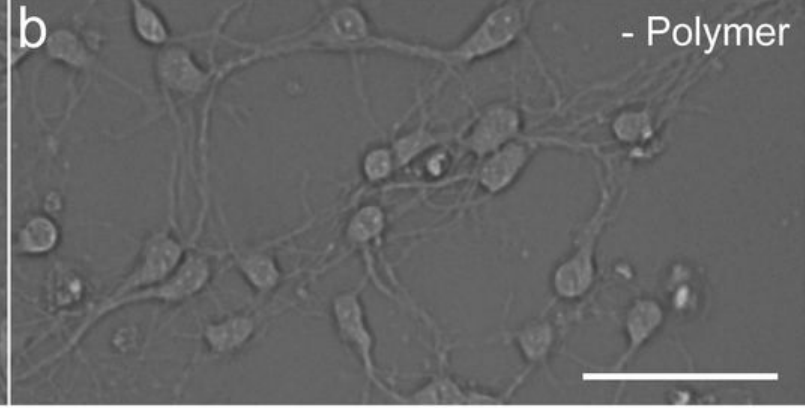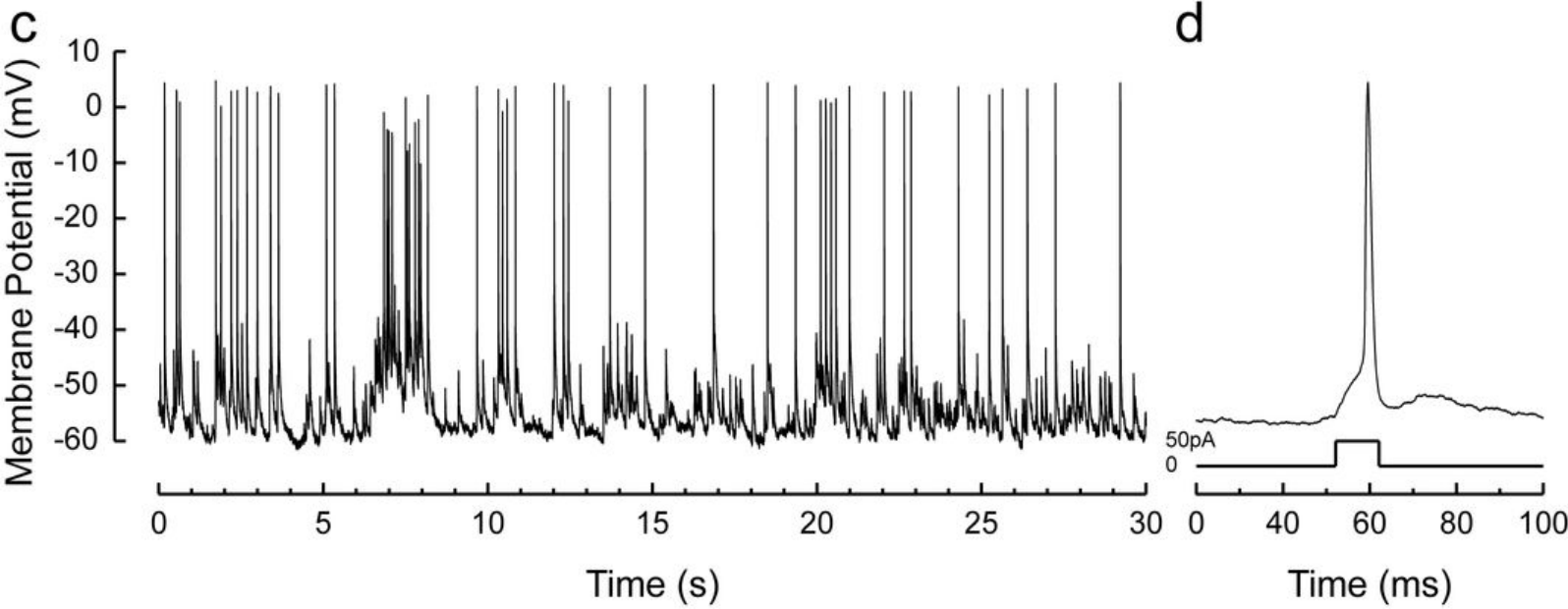

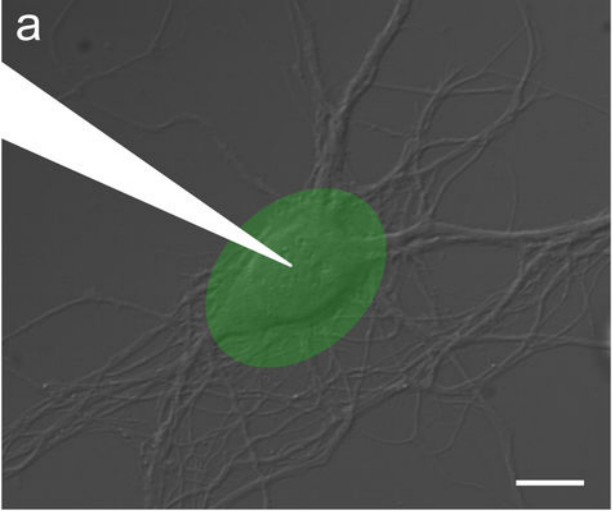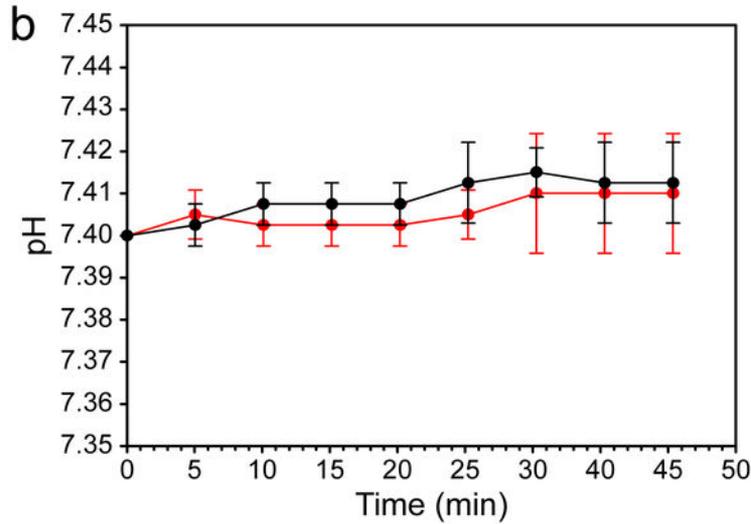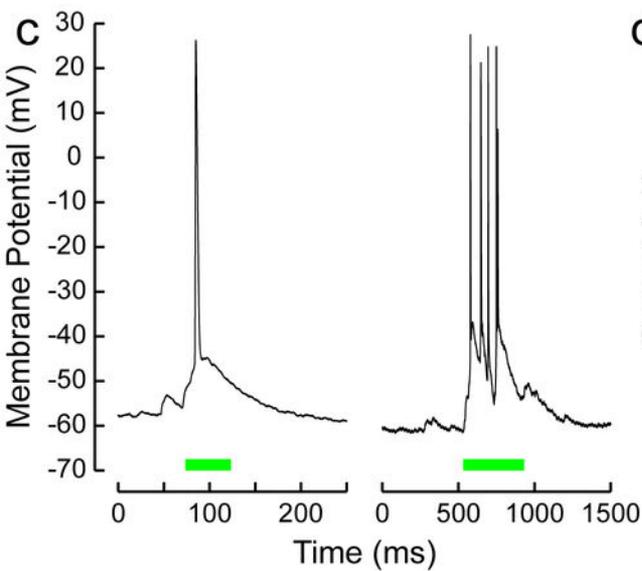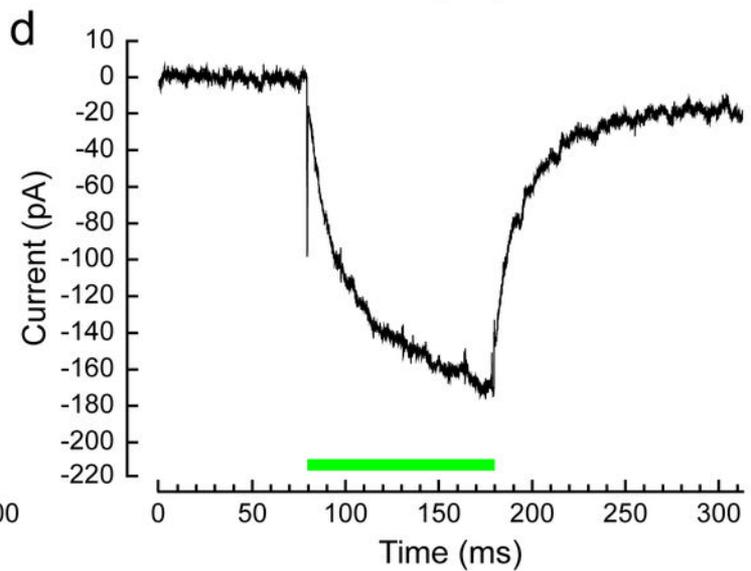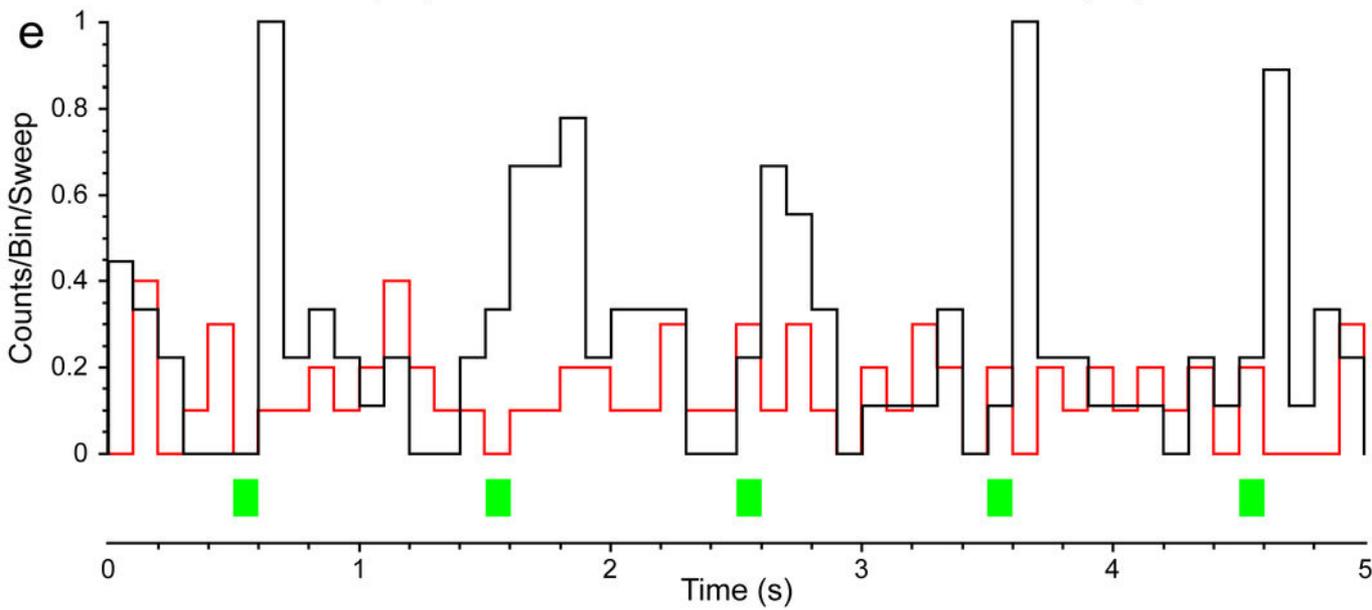